# MOTIVATIONS AND EXPECTATIONS FOR VIRTUAL GIFT-GIVING IN DOUYIN LIVE STREAMS

Huilian Sophie Qiu
Carnegie Mellon University

Daniel Klug
Carnegie Mellon University

**Live Streaming and Virtual Gift-Giving in Social Media and Short-Video Apps**

Social live streaming services (SLSS) (e.g. Twitch, YouTube Live) combine real-time presentation of user-generated content with social networking features that allow streamers and viewers to interact in a co-present social online setting (Zimmer & Scheibe, 2019). Mobile live streaming is now an integrated feature in many social media apps (e.g. Facebook, Instagram, Snapchat) and has recently advanced as a popular form of content creation on entertainment-based short-video apps, especially in the Chinese market (Xu & Ye, 2020) and foremost Douyin, the Chinese version of TikTok.

On Douyin and TikTok, users can create 15 to 60-second videos, usually based on trending sounds, and add hashtags, text elements, and captions to videos. In addition, Douyin promotes its live streaming feature as a way for users to earn money through virtual gifts from viewers (Kaye, Chen, & Zeng, 2020), a common feature and practice in SLSS (Scheibe, Fietkiewicz, Stock, 2016).

Virtual gifts in SLSS are primarily used to reward streamers for their content or performance in a live stream. However, we assume on Douyin social aspects are more crucial for virtual gifting. While attractiveness and humorous appeal of streamers generally affect consumption and virtual gifting (Hou et al., 2019), socialization is also a driving factor in making virtual gifts, and the relationship between streamer and viewer affects the given amount (Yu et al., 2018). Viewers more likely make virtual gifts if streamers show activeness and availability, and showing appreciation for received gifts in a live stream often results in continuous virtual sponsoring (Lee, Choi & Kim, 2019). Seeing streamers receive virtual gifts from other users further encourages gift-giving (Zhu, Yang, & Dai, 2017).

**Three Types of Douyin Live Streams**

We are looking at three types of live streams by young male hosts (= streamers) on Douyin to analyze the motivations of female viewers to watch and make virtual gifts to the hosts and to explore resulting power relations between hosts and viewers.
1) 颜值主播 (y*an zhi zhu bo*, 'good-looking hosts') mostly show their faces and upper body. They chat with viewers, sing, dance, or do a 'PK' (meaning 'player killing'), a 5-minute competition with another host on gaining the most virtual gifts;
2) 声控主播 (s*heng kong zhu bo*, 'vocal hosts') only present their voice to a still image, they chat with viewers, sing, or narrate books;
3) 交友厅 (*jiao you ting*, 'friend-making halls') have several hosts at a time whom viewers can become 'friends' with through chatting but mostly through making virtual gifts for the surplus of getting personal phone calls or being added to the host's WeChat to maintain a personal relationship outside of Douyin.
In all Douyin live streams, viewers can make virtual gifts between ￥0.1 (~$0,015) and ￥6,666 (~$1,017); hosts can create wish lists of virtual gifts for viewers to purchase, in return, they thank viewers in the live stream.

**Method**

We conducted 12 semi-structured interviews with female viewers (ages 18 to 37) of all three live streams who sent virtual gifts to hosts (all live stream viewers can see who sent a virtual gift). We asked why they follow hosts, why and when they decide to make virtual gifts, and what their expectations are from sending virtual gifts. All interviews were conducted in Chinese and translated into analogous English by one of the researchers who is fluent in both languages.

**Findings and Discussion**

Our results confirm that female viewers of male-hosted Douyin live streams mainly use virtual gifts as a social means of appreciation and support for friendship in a parasocial online interaction. Interviewees mostly follow hosts because they are humorous, good-looking, have a nice voice, or are good entertainers, but also because they enjoyed chatting with other fans. Interviewees send virtual gifts to show appreciation for a host's performance because they wanted to make them happy. For example, one interviewee sent virtual gifts up to ￥588 (~$90) to cheer up a host when she felt he was exhausted. They also purchased virtual gifts from wish lists to help a host achieve an assigned virtual gifts task on Douyin. In contrast, one interviewee explained she would not send virtual gifts if a host already received enough from other viewers even when she appreciates his performance. Interviewees rather constantly send smaller amounts than occasional larger amounts but still spend larger amounts overall and prioritize virtual gifts to hosts over buying, for example, leisure articles.

We did not find aspects of power imbalances regarding the monetary or social value of virtual gifts. Most interviewees feel they hold equal positions to the hosts and consider themselves fans and never expected to become friends with a host. However, hosts from 'friend-making halls' live streams are sometimes perceived as friends, yet some viewers of other live streams consider hosts only as a means to find other fans to chat with. Interviewees unanimously said they do not expect anything from the hosts in return for their virtual gifts, yet they appreciate online gestures, like a host blowing a kiss, but do not consider them a privilege. In contrast, our observation of several 'PK' ('player killing') competitions in 'good-looking hosts' Douyin live streams shows viewers send virtual gifts to help their favorite host win the competition which holds implicit expectations.

**Conclusion**

We find these types of Douyin live streams are much more focused on the self-presentation of hosts as personae and actors in parasocial interaction. We are currently further analyzing how hosts use the short-form video format on Douyin to promote live streams as long-form based on self-(re)presentation rather than activities and skills, such as gaming in SLSS.